\documentclass[aip,reprint,english]{revtex4-1}
\usepackage[latin1]{inputenc}
\usepackage{graphics}
\usepackage{epsfig}
\usepackage{amsfonts}
\usepackage{amsmath}
\usepackage{amstext}
\usepackage{amssymb}
\usepackage{graphicx}
\usepackage{natbib}
\usepackage{subfigure}

\usepackage{xcolor}

\usepackage{siunitx} 

\begin{document}

\title{Analyzing critical propagation in a reaction-diffusion-advection model using unstable slow waves}

\author{Frederike Kneer}
\email{fkneer@ni.tu-berlin.de}
\author{Klaus Obermayer}
\affiliation{Department of Software Engineering and Theoretical Computer Science, Technische Universit\"at Berlin, Ernst-Reuter-Platz 7, D-10587 Berlin, Germany}

\author{Markus A. Dahlem}
\affiliation{Department of Physics, Humboldt Universit\"at zu Berlin, Robert-Koch-Platz 4, 10115 Berlin, Berlin, Germany}

\date{\today}

\begin{abstract} \noindent The effect of advection on the critical minimal
  speed of traveling waves is studied. Previous theoretical studies estimated
  the effect on the velocity of stable fast waves and predicted the existence of
  a critical advection strength below which propagating waves are not supported
  anymore. In this paper, the critical advection strength is calculated taking
  into account the unstable slow wave solution. Thereby, theoretical results
  predict, that advection can induce stable wave propagation in the
  non-excitable parameter regime, if the advection strength exceeds a critical
  value. In addition, an analytical expression for the advection-velocity
  relation of the unstable slow wave is derived. Predictions are confirmed
numerically in a two-variable reaction-diffusion model.  \end{abstract}

\maketitle

\section{Introduction}

Traveling waves are basic patterns emerging in excitable media and are observed
in many physical, chemical, and biological systems. In chemical systems,
propagating excitation waves can be found in the Belousov-Zhabotinsky (BZ)
reaction \citep{KEE86,KAP95a}. Many important examples of excitation waves are
found in biological systems, in particular, neuronal systems, such as the
action potential, a wave of electrical depolarization that propagates along the
membrane of a nerve cell axon with constant shape and velocity \citep{HOD52},
or spreading depression (SD), a wave of sustained cell and tissue
depolarization caused by a massive release of Gibbs free energy that propagates
through gray matter tissue \cite{LEA44,DRE11}.  Besides, intracellular waves of
calcium have been observed \citep{CAM93, FAL99}.  In physical systems, a large
variety of spatiotemporal patterns has been shown to occur during the oxidation
of CO on a Pt(110) surface \citep{JAK90,BET04a,BAE94}. 

As a model for these traveling waves, we consider excitable media of
activator-inhibitor type.  This macroscopic description is used to study the
generic behavior of traveling waves in reaction-diffusion-advection systems.
The spatial coupling within the medium is primarily  given by diffusion, while
advection is introduced in either of two ways.  First, external forcing can
lead to advection which changes the excitation properties of the unforced
reaction-diffusion system.  In this case, the advection term models the mean
flow, for instance, of ions driven by an externally applied constant electrical
field \citep{ISI:A1986D719600004,GomezGesteira1997389}. This case has been studied in the chemical BZ reaction
\citep{STE92,Chomaz_1992_PRL} and in
some preliminary studies in cortical SD \cite{GRA56}. Second, in
two-dimensional reaction-diffusion media, a small curvature of a wave front can
also formally lead to an advection term under some approximations resulting in a
reduced reaction-diffusion-advection description in one dimension \citep{ZYK84,GomezGesteira1997389,TYS88}.  Front curvature effects have been observed in
the BZ reaction \cite{foerster1988curvature,STE93a}. Furthermore, drifting pulses that
form via an advection instability in a reaction-diffusion medium with
differential advection have been analyzed \citep{yochelis2010dri} and critical
properties of traveling waves affected by advection have been discussed
\citep{ZYK84,DAV02,GomezGesteira1997389}.

It has been shown, that advection can have destructive and constructive effects
on traveling waves, namely, slowing them down and even abolish them at a
critical speed, and accelerating them and even facilitate propagation of
traveling waves in the parameter regime in which the system without advection
is non-excitable, respectively.   Here we
investigate in particular the latter non-excitable regime, which without advection does not support traveling waves.
In this regime, the current analytical
approximation fails. We provide an extended analytical approximation and
compare our results also with numerically simulations.

\section{FitzHugh-Nagumo in co-moving frame and with advection}
\label{sec:FHN}

\subsection{FitzHugh-Nagumo dynamics}

Let us firstly consider excitable media of activator-inhibitor type
 in one spatial dimension with diffusion,
\begin{eqnarray} 
  \frac{\partial u}{\partial t} &=& f(u,v) +D_u \frac{\partial^2 u}{\partial x^2},\label{eq:FHNohneAu}\\ 
  \frac{\partial v}{\partial t} &=& \varepsilon g(u,v) + D_v \frac{\partial^2 u}{\partial x^2}.\label{eq:FHNohneAv} 
\end{eqnarray} 

This system has two variables $u(x,t)$ and $v(x,t)$ called activator and
inhibitor, respectively, that depend on time $t$ and space $x$. The parameters
$D_u$ and $D_v$ are the diffusion coefficients of activator $u$ and inhibitor
$v$, respectively. The parameter $\varepsilon$ is the time scale ratio between
$u$ and $v$. 

Next, we specify the activator rate function $f(u,v)$ and inhibitor rate
function $g(u,v)$ as FitzHugh-Nagumo dynamics \citep{BON48,FIT61,NAG62}, that
is, $f(u,v)=3u-u^3-v$ and  $g(u,v)=(u+ \beta + \gamma v)$. Note, that in the
most general case of FitzHugh-Nagumo systems---defined as
$f(u,v)$ having a cubic nonlinearity in the first argument $u$ and being linear
otherwise, in particular, $g(u,v)$ is linear---there are only three free parameters:
$\varepsilon$, $\beta$, and $\gamma$. With diffusion, only one more free parameter is introduced, because
one of the two diffusion coefficients $D_u$
and $D_v$ can be set to unity by scaling space.

FitzHugh-Nagumo dynamics is chosen, as it provides a mathematically tractable excitable
medium of activator-inhibitor type and we further simplify this system to
obtain only two free parameter as follows. Inhibitor diffusion is assumed to be
slow and hence negligible, i.e., $D_v=0$. In the remainder, we refer to $D_u$
as $D$ and note that formally, it is not a free parameter anymore as it can be
set to unity by scaling $x$ accordingly. Moreover, we chose to set $\gamma =0
$. These simplifications are further discussed in Sec.~\ref{sec:conclusion}.

With only the two parameters $\varepsilon$ and $\beta$ left, the influence of
an additional advection term is more easy to illustrate and also the suitable
regime of $\beta$ can readily be seen. The parameter $\varepsilon$ has
to be chosen, in any case, much smaller than unity, because only slow
inhibitor kinetics render dynamics excitable.  In the local FitzHugh-Nagumo system
($D_u\!=\!D_v\!=\!0$) and at any arbitrary position $x_0$, the parameter
$\beta$ determines whether the dynamics at $x_0$ is in the excitable regime
($\beta\!>\!1$) or exhibits self-sustained periodic oscillations
($\beta\!<\!1)$. In the following, we only consider the excitable regime, which 
is in parameter space near the oscillatory regime.

\subsection{Traveling waves and co-moving coordinate frame}

\begin{figure}
\includegraphics[width=\columnwidth]{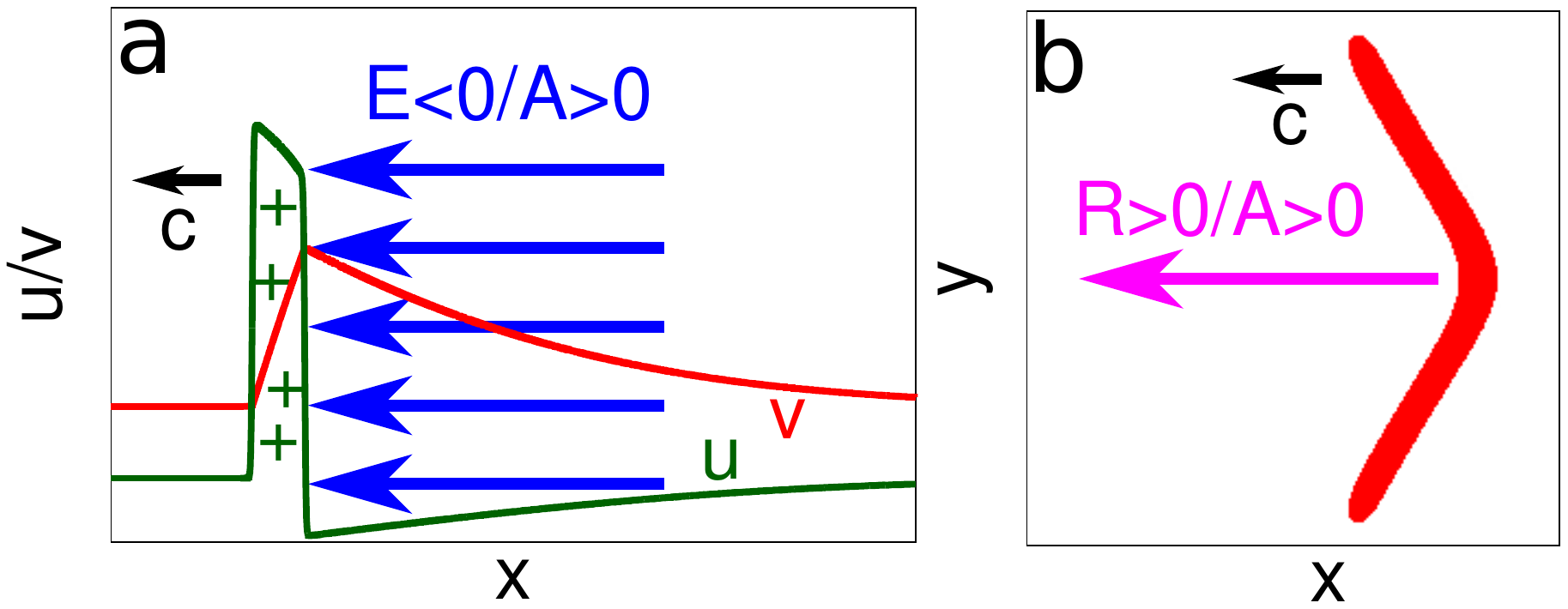}
\caption{Illustration of (a) activator and inhibitor profiles propagating in an external electrical field and (b) two-dimensional V-shaped pattern (top view, red indicates the area with $u>0$).}
  \label{fig:pulseCurve}
\end{figure}

Next, we consider traveling waves, i.e., solutions of
Eqs.~(\ref{eq:FHNohneAu})-(\ref{eq:FHNohneAv}) with a constant propagation
velocity $c$ and unaltered wave profile $u(x,t)=u(\xi)$, $v(x,t)=v(\xi)$, that
is, a stationary profile in the co-moving coordinate $\xi=x+ct$.  Without loss
of generality, we only consider waves propagating in negative $x$-direction,
see Fig.\ \ref{fig:pulseCurve}a.\\

Traveling waves are stationary profiles in co-moving coordinate
frames. To find these, Eqs.~(\ref{eq:FHNohneAu})-(\ref{eq:FHNohneAv}) then can be
transformed to   \begin{eqnarray} c\frac{\partial u}{\partial\xi} &=&
3u-u^3-v+D\frac{\partial^2 u}{\partial\xi^2},\label{eq:FHNcomovingohneAu}\\
c\frac{\partial v}{\partial\xi} &=& \varepsilon(u+\beta).
\label{eq:FHNcomovingohneAv} \end{eqnarray} An advection term, added to
Eq.~(\ref{eq:FHNohneAu}) or
Eq.~(\ref{eq:FHNcomovingohneAu}), may arise
through different mechanisms. 

\subsection{Advection}

Let us only briefly mention the quantities and how they relate formally to an
advection term in an 1D approximation of curved RD fronts in spatially
two-dimensional media\citep{ZYK84}. Propagating slightly curved wave fronts $(R\ll L)$,
where $L$ is the width of the rising front, can be approximated by 
\begin{eqnarray}
 c(A)\frac{\partial u}{\partial\xi} &=& 3u-u^3-v+D\frac{\partial^2 u}{\partial\xi^2}+A\frac{\partial u}{\partial\xi},\label{eq:FHNcomovingRu}\\
c(A)\frac{\partial v}{\partial\xi} &=& \varepsilon(u+\beta), 
\label{eq:FHNcomovingRv}
\end{eqnarray}
with $A=\frac{D}{R}$, where $R$ is the curvature radius of the front. For a
detailed derivation, see Ref.~\citep{ZYK84}. The term $A\frac{\partial
u}{\partial\xi}$ is called advection term. 

As it is not readily obvious, we will also briefly derive that basically the
same set of equations, i.e.,
Eqs.~(\ref{eq:FHNcomovingRu})-(\ref{eq:FHNcomovingRv}), can be obtained, if one
considers advection due to a constant external driving force.  Both, activator
$u$ and inhibitor $v$ can be associated with particles of different mobilities
$m_u$ and $m_v$. It seems that we can also neglect $m_v$, because we already
assumed inhibitor diffusion to be negligible and the diffusion coefficient is
related to the mobility through the Einstein relation $D=mkT$, where $k$ is
Boltzmann's constant, and $T$ the absolute temperature.  Note, however, that we
have to consider the electrical mobility $\mu$, which is the mobility $m$ times
the charge $q$ of the particle. For ions or charged macromolecules, the charge
$q$ is the valence number $z$ times the elementary charge $e$ of the electron,
thus, $\mu=mze$. Therefore, the absolute value of the quotient of the
electrical mobilities $|\mu_v/\mu_u|$ is not necessarily much smaller than
unity, even if $m_v/m_u\ll1$. Since a large valence number $z$ is only found in
large charged macromolecules, which indeed have a much smaller mobility $m$, an
advection term in the inhibitor equation despite the fact that we set the
diffusion to zero is a reasonable assumption.

Particle motion
could then be affected by a homogeneous external field (e.g. an electrical
field of strength $E$), which is applied parallel to the propagation direction,
and Eqs.~(\ref{eq:FHNcomovingohneAu})-(\ref{eq:FHNcomovingohneAv}) read
\begin{eqnarray}
 c\frac{\partial u}{\partial\xi} &=& 3u-u^3-v+D\frac{\partial^2 u}{\partial\xi^2}+\mu_uF \frac{\partial u}{\partial\xi},\label{eq:FHNcomovingmobilityu}\\
c\frac{\partial v}{\partial\xi} &=& \varepsilon(u+\beta)+\mu_vF\frac{\partial v}{\partial\xi}, 
\label{eq:FHNcomovingmobilityv}
\end{eqnarray}
where $F$ is the strength of the field and $zE=-F$ with the valence $z$ of the ion.\\
Changing the velocity of the co-moving frame to $\tilde c$,
\begin{eqnarray}
 \tilde c&=&c-\mu_vF.
\end{eqnarray}
One can interpret this system in the co-moving frame with $\tilde c$ as a system with advection only in the activator with advection strength $A$, see Fig.\ \ref{fig:pulseCurve}a. For $\tilde c=c(A)$, this yields
\begin{eqnarray}
 c(A)\frac{\partial u}{\partial\xi} &=& 3u-u^3-v+D\frac{\partial^2 u}{\partial\xi^2}+A\frac{\partial u}{\partial\xi}\label{eq:FHNcomovingu},\\
c(A)\frac{\partial v}{\partial\xi} &=& \varepsilon(u+\beta)\label{eq:FHNcomovingv}, 
\end{eqnarray}
where $\xi=x-(c-\mu_vF)t$ and $A=F(\mu_u-\mu_v)$. The first and the second mechanism now are described by the same equation, as Eqs.~(\ref{eq:FHNcomovingu})-(\ref{eq:FHNcomovingv}) and Eqs.~(\ref{eq:FHNcomovingRu})-(\ref{eq:FHNcomovingRv}) are the same.
In stationary coordinates, this reads
\begin{eqnarray}
 \frac{\partial u}{\partial t} &=& 3u-u^3-v+D\frac{\partial^2 u}{\partial x^2}+A\frac{\partial u}{\partial x},\label{eq:FHNu}\\
\frac{\partial v}{\partial t} &=& \varepsilon(u+\beta).
\label{eq:FHNv}
\end{eqnarray}
For $c(A)>0$ (propagation in negative $x$-direction), $A>0$ can be interpreted as an approximation to small positive curvature of a wave front propagationg in a spatial 2D medium, that e.g.\ exhibit so-called V-shaped pattern \citep{Brazhnik199540}, see Fig.~\ref{fig:pulseCurve}b. Interpreting the activator variable $u$ and the inhibitor variable $v$ as the concentration of different charged ions, $A>0$ corresponds to a constant electrical field externally applied parallel to the propagation direction. For example, activator $u$ being positive charged ions and inhibitot $v$ being noncharged, $A>0$ corresponds to an electrical field that has the same direction as the propagation direction, see Fig.~\ref{fig:pulseCurve}a.

\section{Theory}

In this section, we derive an approximation for the critical velocity and the
corresponding critical advection strength Sec.~\ref{sec:crvelandcradv}. 
To this end, we first define the propagation boundary Sec.~\ref{sec:pb} and then derive the
advection-velocity relation for unstable waves in Sec.~\ref{sec:advelslowwave}.

\subsection{Propagation boundary}
\label{sec:pb}

FitzHugh-Nagumo system without advection (Eqs.~(\ref{eq:FHNohneAu})-(\ref{eq:FHNohneAv})) ($1<\beta<\sqrt{3}$ and $\varepsilon$ sufficiently small) has a stable fast wave solution and an unstable slow wave solution which correspond to homoclinic orbits of the related ODE problem (Eqs.~(\ref{eq:FHNcomovingohneAu})-(\ref{eq:FHNcomovingohneAv})), see Ref.\citep{Krupa199749}. There exists a critical line $\partial P$ in the $(\varepsilon,\beta)$ space, at which the fast wave branch collides with the slow wave branch.  For values of $\beta$ and $\varepsilon$ above this critical line, propagation of traveling waves cannot be obtained. These properties carry over to the case of finite advection strength $A$. Thus it is reasonable to take into account the slow wave solution when calculating the critical properties, i.e. the critical surface in the $(\varepsilon,\beta,A)$ space, which separates the excitable and the non-excitable parameter regime and a critical velocity $c_{cr}$ depending on advection $A$.

\begin{figure}
\includegraphics[width=\columnwidth]{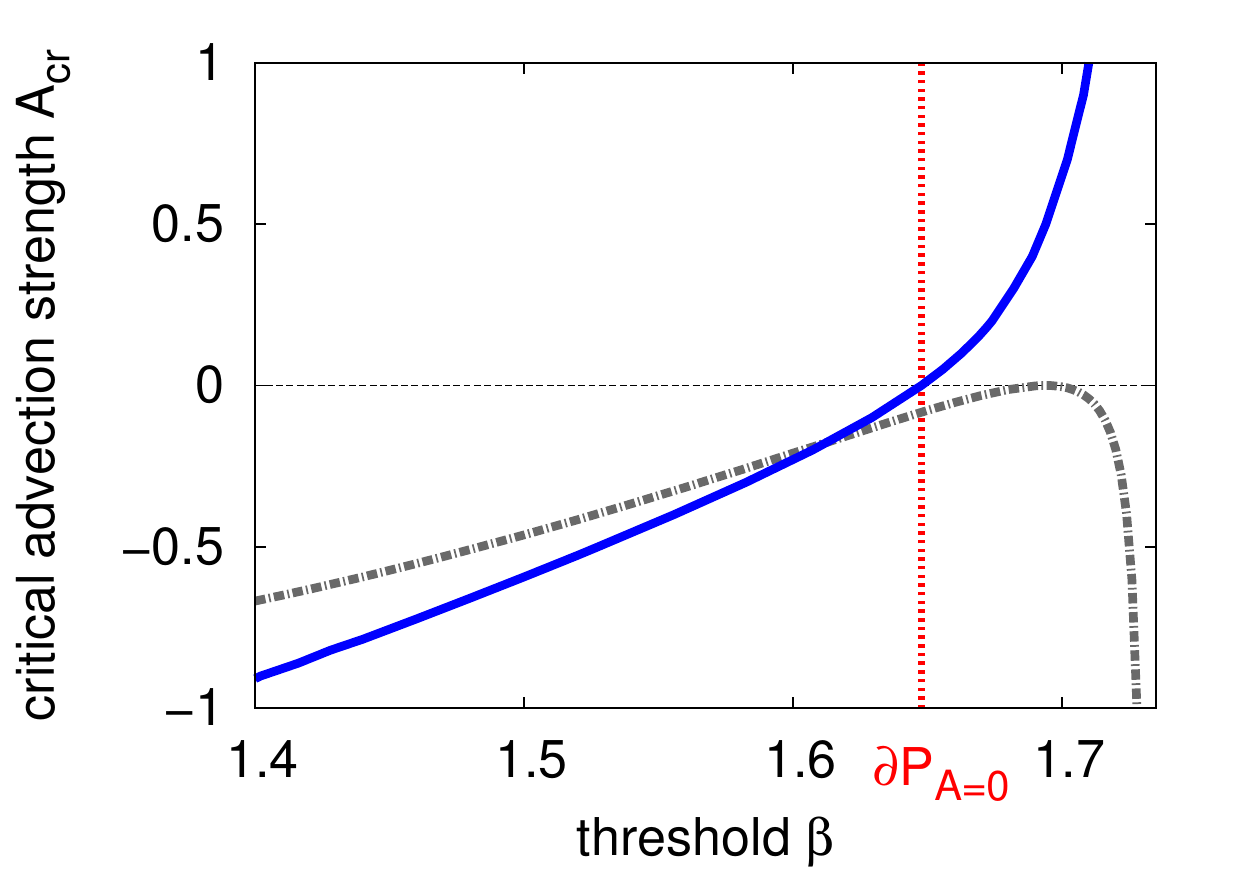}
 \caption{Critical advection strength $A_{cr}$ as a function of threshold size $\beta$. The grey dashed line shows the results from Eq.(\ref{eq:acreikonal}), which was derived from the nonlinear Eikonal equation. The blue solid line shows the results from Eqs.~(\ref{eq:FHNu})-(\ref{eq:FHNv}) computed with $AUTO$ by continuating homoclinic solutions; the propagation boundary $\partial P_{A=0}$ is computed from Eqs.~(\ref{eq:FHNohneAu})-(\ref{eq:FHNohneAv}). ($\varepsilon=0.022$ in all cases.)  The blue solid line separates the excitable from the non-excitable parameter regime.}
  \label{fig:Acreik}
\end{figure}

The nonlinear Eikonal equation \citep{ZYK84} provides a good approximation for the advection-velocity relation of the fast wave solution, if the wave speed is decelerated ($A<0$). The nonlinear Eikonal equation has also been used to calculate a critical velocity and a critical advection strength \citep{ZYK84}.
The critical advection strength $A_{cr}$ derived from the nonlinear Eikonal equation provides a good approximation for the critical advection strength $A_{cr}$ (see Appendix) needed for loss of excitability in the parameter regime $\beta<\partial P_{A=0}$ (see Fig.\ref{fig:Acreik}). Under the influence of advection $A<0$, the propagation boundary is shifted to smaller threshold values $\beta$.

Numerical calculations (see Sec.~\ref{sec:results}) show that positive advection $A>0$ induces stable wave propagation in the parameter regime $\beta>\partial P_{A=0}$, the propagation boundary is shifted to larger threshold values $\beta$.
In this parameter range, theory strongly deviates from numerical calculations, thus this behaviour is not explained by the nonlinear Eikonal equation.\\

\subsection{Advection-velocity relation for the fast and slow wave solution}
\label{sec:advelslowwave}

In this section, the advection-velocity relation of the slow wave is derived in the same way as the known advection-velocity relation for the fast wave.  
Rewriting Eqs.~(\ref{eq:FHNcomovingu})-(\ref{eq:FHNcomovingv}) (see Ref.\citep{ ZYK84}), 
\begin{eqnarray}
 (c(A)-A)\frac{\partial u}{\partial\xi} &=& 3u-u^3-v+D\frac{\partial^2 u}{\partial\xi^2},\label{eq:FHNansatzEiku}\\
c(A)\frac{\partial v}{\partial\xi} &=& \varepsilon(u+\beta). 
\label{eq:FHNansatzEikv}
\end{eqnarray}
and introducing $c^*$ and $\varepsilon^*$ 
\begin{eqnarray}
 c^*&=&c(A)-A,\label{eq:cstar}\\
 \varepsilon^*&=&\varepsilon\frac{c^*}{c(A)},
 \label{eq:epsstar}
\end{eqnarray}
yields
\begin{eqnarray}
 c^*\frac{\partial u}{\partial\xi} &=& 3u-u^3-v+D\frac{\partial^2 u}{\partial\xi^2},\label{eq:FHNstaru}\\
c^*\frac{\partial v}{\partial\xi} &=& \varepsilon^*(u+\beta), 
\label{eq:FHNstarv}
\end{eqnarray}
which has the same form as the FitzHugh-Nagumo model without advection (Eqs.~(\ref{eq:FHNcomovingohneAu})-(\ref{eq:FHNcomovingohneAv})). Thus $c^*$ has the same dependency on $\varepsilon^*$ and $\beta$ as the propagation velocity $c|_{A=0}$  (see Eqs.~(\ref{eq:FHNcomovingohneAu})-(\ref{eq:FHNcomovingohneAv})) on $\varepsilon$ and $\beta$. The velocity $c|_{A=0}$ for the fast and the slow wave can then approximately be calculated using a singular perturbation theory \citep{Casten123456}. 
The propagation velocity of the fast, $c^f|_{A=0}$, and the slow, $c^s|_{A=0}$, wave is then obtained of
\begin{eqnarray}
 c^f|_{A=0}&=&c_0+\varepsilon c_1^f,\label{eq:cfcsf}\\
 c^s|_{A=0}&=&\sqrt{\varepsilon} c_1^s.
 \label{eq:cfcss}
\end{eqnarray}
The expressions for $c_0$, $c_1^f$ and $c_1^s$ are provided in the Appendix.\\
For $c^*$ Eqs.~(\ref{eq:FHNstaru})-(\ref{eq:FHNstarv}) we, therefore, obtain the expressions
\begin{eqnarray}
 c^{f*}&=&c_0+\varepsilon^* c_1^f,\label{eq:cfs}\\
 c^{s*}&=&\sqrt{\varepsilon^*} c_1^s.\label{eq:css}.
\end{eqnarray}
Inserting $c^*=c(A)-A$ and $\varepsilon^*=\varepsilon\frac{c^*}{c(A)}=\varepsilon\frac{c(A)-A}{c(A)}$ (see Eqs.~(\ref{eq:cstar})-(\ref{eq:epsstar})), we obtain
\begin{eqnarray}
 c^f(A)-A&=&c_0+\varepsilon\frac{c^f(A)-A}{c^f(A)}c_1^f,\\
 c^s(A)-A&=&\sqrt{\varepsilon\frac{c^s(A)-A}{c^s(A)}}c_1^s.
 \label{eq:cfacsa}
\end{eqnarray}
Solving for $c^f(A)$, we obtain the so-called nonlinear Eikonal equation
\begin{eqnarray}
 c^f_{\pm }(A)&=&\frac{1}{2}((A+c_0+\varepsilon c_1)\pm\sqrt{(A+c_0+\varepsilon c_1)^2-4\varepsilon A c_1}),\nonumber\\
 \label{eq:cf12}
\end{eqnarray}
where $c^f_+(A)$ is the valid advection-velocity relation, because $c_+^f|_{A=0}=c_++\varepsilon c_1^f$, see \citep{ ZYK84}.\\
Solving Eq.(\ref{eq:cfacsa}) for $c^s(A)$, we obtain
\begin{eqnarray}
 c^s_{\pm }(A)&=&\frac{1}{2}(A\pm\sqrt{A^2+4\varepsilon c_1^{s2}}),\nonumber\\
 c^s_3(A)&=&A.
 \label{eq:cs12}
\end{eqnarray}
The valid advection-velocity relation for the slow wave (with $c^s(A)>0$)  is $c^s_+(A)$, because $c^s|_{A=0}\equiv\sqrt{\varepsilon}c_1^s$.

\subsection{Critical velocity and critical advection strength}
\label{sec:crvelandcradv}

Here, the critical velocity $c_{cr}(A_{cr})$, which exhibits a traveling wave at the connection of the fast wave and the slow wave branch affected by a critical advection of strength $A_{cr}$, is calculated. Also an expression for $A_{cr}$ is captured by this calculations.\\
In a FitzHugh-Nagumo model without advection (Eqs.~(\ref{eq:FHNohneAu})-(\ref{eq:FHNohneAv})), there exists a critical line in the $(\varepsilon,\beta)$ parameter space, above which wave propagation is impossible. At the critical time scale ratio $\varepsilon_{cr}$, the single homoclinic solution of Eqs.~(\ref{eq:FHNcomovingohneAu})-(\ref{eq:FHNcomovingohneAv}) corresponds to the connection between the fast wave branch and the slow wave branch, and the propagation velocity of the fast wave is minimal ($c_{cr}|_{A=0}$).\\
The critical time scale ratio $\varepsilon_{cr}$ as a function of $\beta$ can be approximated by solving $c^s|_{A=0}=c^f|_{A=0}$ for $\varepsilon_{cr}$, where $c^f|_{A=0}$ and $c^s|_{A=0}$ are calculated using singular perturbation theory (Eqs.~(\ref{eq:cfcsf})-(\ref{eq:cfcss})). This yields
\begin{eqnarray}
\varepsilon_{cr}^{\pm }(\beta)&=&\frac{-2c_0c_1^f+c_1^{s2}\pm\sqrt{-4c_0c_1^fc_1^{s2}+c_1^{s4}}}{2c_1^{f2}},
\label{eq:evonb}
\end{eqnarray}
where $\varepsilon_{cr}^-<\varepsilon_{cr}^+$ and thus $\varepsilon_{cr}=\varepsilon_{cr}^-$, compare Sec.(\ref{sec:results}). For the critical velocity $c_{cr}|_{A=0}$ as a function of $\beta$ we then obtain of Eqs.~(\ref{eq:cfcsf})-(\ref{eq:cfcss})
\begin{eqnarray}
 c_{cr}|_{A=0}&=&c_0+\varepsilon_{cr}c_1^f=\sqrt{\varepsilon_{cr}}c_1^s.
 \label{eq:ccr}
\end{eqnarray}

Advection changes the critical velocity. To obtain an analytical expression for $c_{cr}(A_{cr})$, we again start from Eqs.~(\ref{eq:FHNstaru})-(\ref{eq:FHNstarv}), which has the same form as FitzHugh-Nagumo model without advection Eqs.~(\ref{eq:FHNcomovingohneAu})-(\ref{eq:FHNcomovingohneAv}). Substituting $c^*$ for $c_{cr}|_{A=0}$ and $\varepsilon^*$ for $\varepsilon_{cr}$, the homoclinic solution of Eqs.~(\ref{eq:FHNstaru})-(\ref{eq:FHNstarv}) ceases to exist at the connection between the fast wave branch and the slow wave branch. Thus, the critical velocity $c_{cr}(A_{cr})$ in systems affected by advection can be derived from Eqs.~(\ref{eq:cstar})-(\ref{eq:epsstar}) by setting $c^*=c_{cr}|_{A=0}$ and $\varepsilon^*=\varepsilon_{cr}$. With $c^*=c(A)-A$  and $\varepsilon^*=\varepsilon\frac{c^*}{c(A)}$ it follows, that
\begin{eqnarray}
 c_{cr}|_{A=0}&=&c_{cr}(A_{cr})-A_{cr},\label{eq:ccracr} \\
 \varepsilon_{cr}&=&\varepsilon\frac{c_{cr}|_{A=0}}{c_{cr}(A_{cr})},
 \label{eq:ccrepscr}
\end{eqnarray}
where $c_{cr}|_{A=0}$ is the minimal propagation velocity of the fast wave for $A=0$ (Eq.(\ref{eq:ccr})) and $c_{cr}(A_{cr})$ is the minimal propagation velocity of the fast wave, that can be achieved by influencing the system with critical advection $A_{cr}$.\\
Solving Eq.(\ref{eq:ccrepscr}) for $c_{cr}(A_{cr})$ and Eq.(\ref{eq:ccracr}) for $A_{cr}$, we finally obtain
\begin{eqnarray}
 c_{cr}(A_{cr})&=&\frac{\varepsilon}{\varepsilon_{cr}}c_{cr}|_{A=0},
 \label{eq:cmin}\\
 A_{cr}=c_{cr}(A_{cr})-c_{cr}|_{A=0}&=&c_{cr}|_{A=0}(\frac{\varepsilon}{\varepsilon_{cr}}-1).
 \label{eq:Acr}
\end{eqnarray}
Be aware that $c_{cr}|_{A=0}$ Eq.(\ref{eq:ccr}) as well as $\varepsilon_{cr}$ Eq.(\ref{eq:evonb}) are fully determined by $\beta$. Thus Eq.(\ref{eq:Acr}) is an approximation for the critical surface in the $(\varepsilon,\beta,A)$ space, above which propagating waves are not supported. As a function of $A$ and $\beta$ it reads
\begin{eqnarray}
 \varepsilon&=&\frac{(A+c_{cr}|_{A=0})\varepsilon_{cr}}{c_{cr}|_{A=0}}.
 \label{eq:criticalsurface}
\end{eqnarray}
For values of $\varepsilon$ above this critical surface, wave propagation is impossible.

\section{Numerical validation}
\label{sec:results}
\begin{figure}[!]
\includegraphics[width=.45\textwidth]{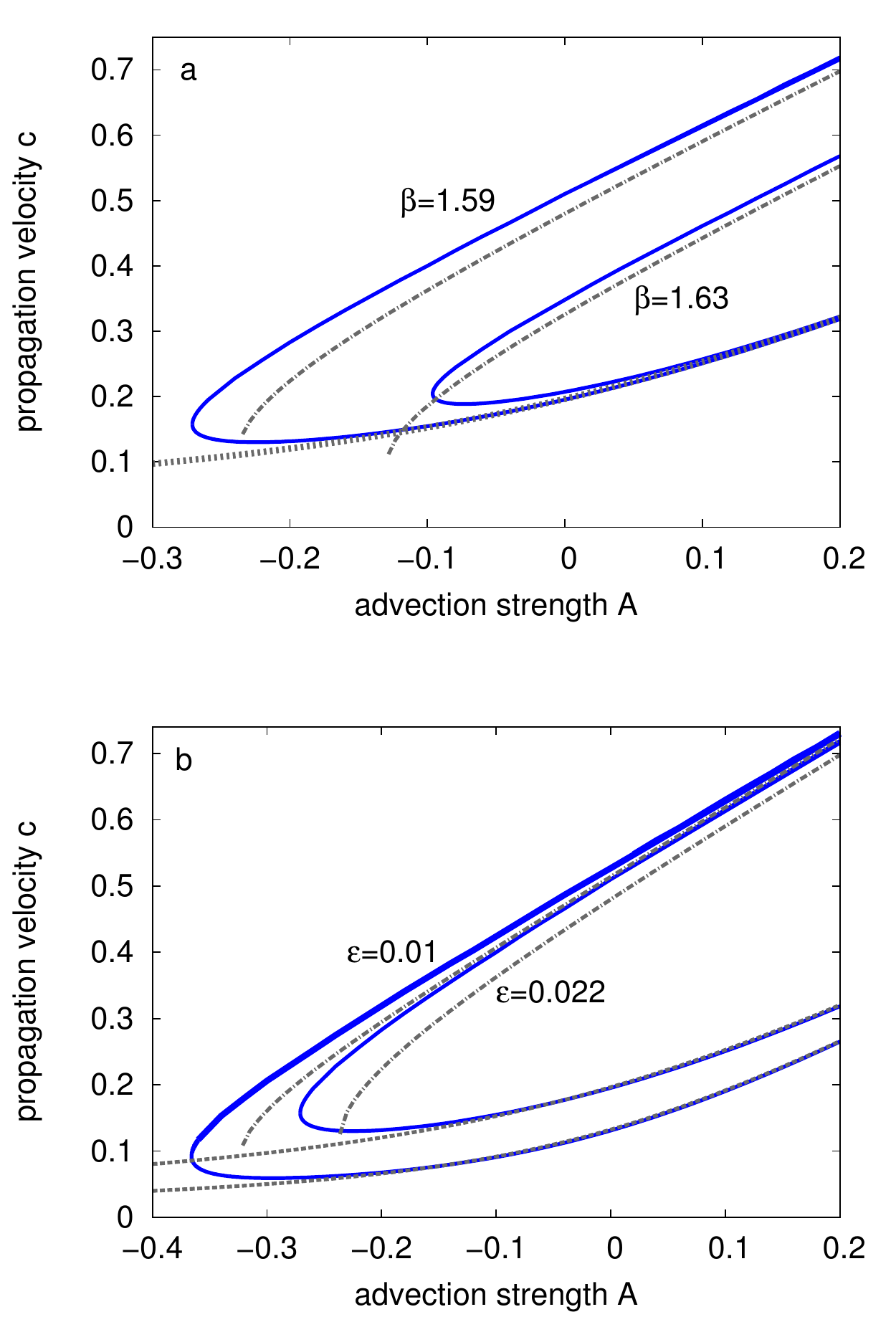}
  \caption{Propagation velocity $c$ as a function of advection strength $A$. The grey dashed-dotted lines show the velocity of the fast wave calculated from the nonlinear Eikonal equation ($c_+^f(A)$ of Eq.(\ref{eq:cf12})). The grey dashed lines show the slow wave velocity derived from Eq.(\ref{eq:cs12}) ($c_+^s(A)$). The blue solid lines show the results numerically computed from Eqs.~(\ref{eq:FHNu})-(\ref{eq:FHNv}). a) $\varepsilon=0.022$, b) $\beta=1.59$.}
 \label{fig:cfcs}
\end{figure}

Fig.(\ref{fig:cfcs}) shows the propagation velocity of the fast, $c^f(A)$, and the slow, $c^s(A)$, waves as a function of advection strength $A$ for different values of $\beta$ (Fig.(\ref{fig:cfcs})(a)) and $\varepsilon$ (Fig.(\ref{fig:cfcs})(b)). The analytical advection-velocity relation for the slow wave Eq.(\ref{eq:cs12}) as well as the nonlinear Eikonal equation Eq.(\ref{eq:cf12}), which provides the advection-velocity relation for the fast wave, are compared with numerical results directly obtained from Eqs.~(\ref{eq:FHNu})-(\ref{eq:FHNv}). We find, that the results from the nonlinear Eikonal equation lie below the numerical results in each case. This is in accordance with the propagation velocity of the fast wave solution $c^f|_{A=0}$ calculated with the singular perturbation theory, which lies below the exact results in the whole parameter regime (except for some parameter values close to the saddle-node bifurcation point, where perturbation theory is less accurate). The larger $\varepsilon$ is, the 
larger is the 
deviation, as the calculations depend on small values of $\varepsilon$.\\
In addition, we find that the advection-velocity relation for the slow wave is more accurate compared to the nonlinear Eikonal equation.
This again is in accordance with the singular perturbation theory, which in the shown parameter regime provides more accurate results for the slow wave velocity $c^s|_{A=0}$ than for the fast wave velocity $c^f|_{A=0}$. Close to the point where the fast wave branch and the slow wave branch meet, the advection-velocity relation for the slow wave deviates more strongly from numerical results, because perturbation theory does not capture the bifurcation behaviour.\\
Furthermore, the analytical results become less accurate for large negative adevction $A<0$, because the results are obtained using a singular perturbation theory depending on small changes in $\varepsilon^*$, see Sec.\ref{sec:advelslowwave}, and $\varepsilon^*=\varepsilon(1-\frac{A}{c(A)})$ Eqs.~(\ref{eq:cstar})-(\ref{eq:epsstar}) increases for increasing absolute value of advection strength $A<0$.\\

\begin{figure}
\includegraphics[width=\columnwidth]{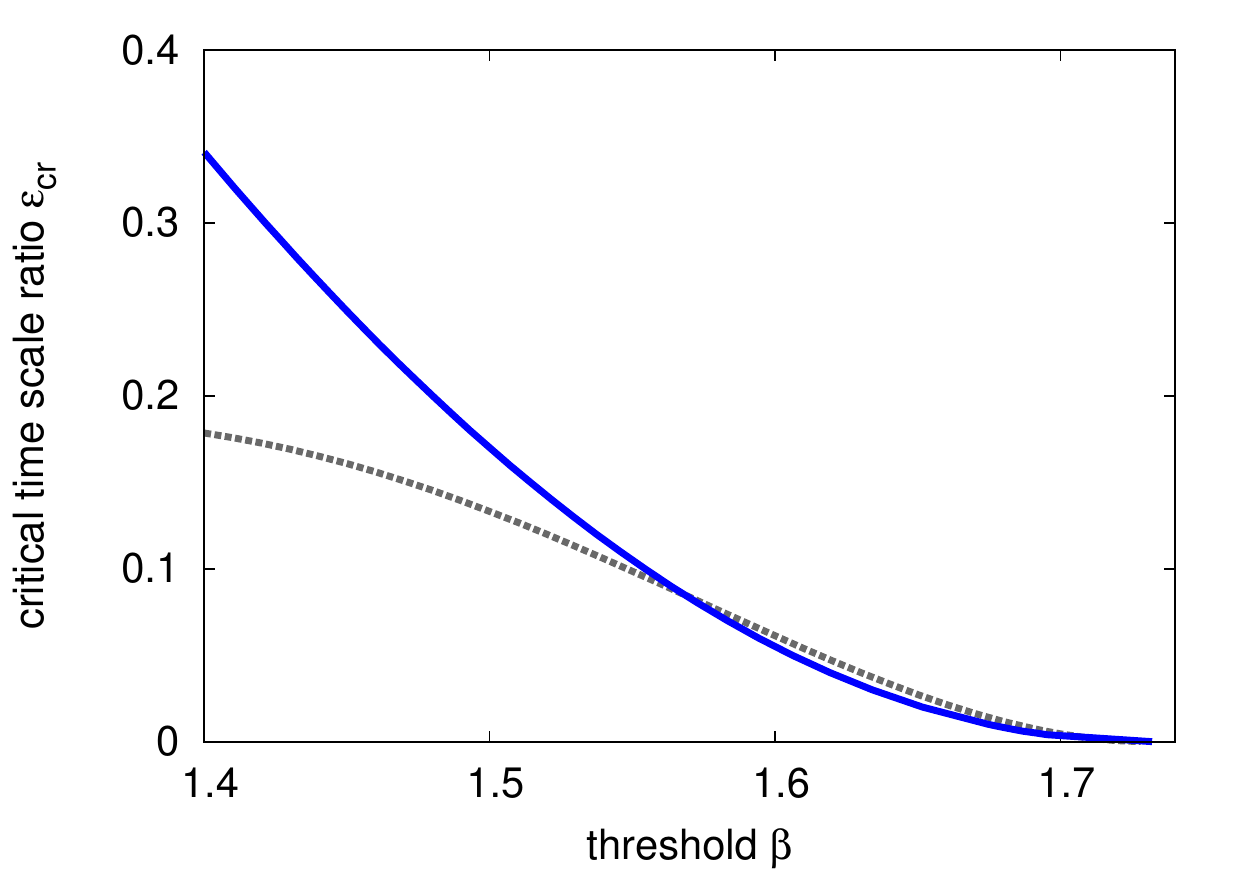}
 \caption{Critical time scale ratio $\varepsilon_{cr}$ as a function of threshold $\beta$. The grey dashed line shows the results derived from Eq.(\ref{eq:evonb}); the blue solid line shows the results numerically computed results from Eqs.~(\ref{eq:FHNohneAu})-(\ref{eq:FHNohneAv}). $A=0$ in each case.}
  \label{fig:evonbmy}
\end{figure}

Fig.\ref{fig:evonbmy} shows the critical time scale ratio $\varepsilon_{cr}$, a property of FitzHugh-Nagumo system without advection Eqs.~(\ref{eq:FHNohneAu})-(\ref{eq:FHNohneAv}), see Sec.\ref{sec:crvelandcradv}, as a function of threshold $\beta$. For $\varepsilon>\varepsilon_{cr}$, the system is non-excitable. The analytical results from Eq.(\ref{eq:evonb}) are compared to numerical results directly obtained from Eqs.~(\ref{eq:FHNohneAu})-(\ref{eq:FHNohneAv}). We find, that for $\varepsilon<0.1$, Eq.(\ref{eq:evonb}) provides a good approximation for the critical time scale ratio $\varepsilon_{cr}$, the absolute error is less than $0.01$. For larger values of $\varepsilon$, the deviation increases, because Eq.(\ref{eq:evonb}) is based on a singular perturbation theory depending on small values of $\varepsilon$.\\

\begin{figure}
\includegraphics[width=\columnwidth]{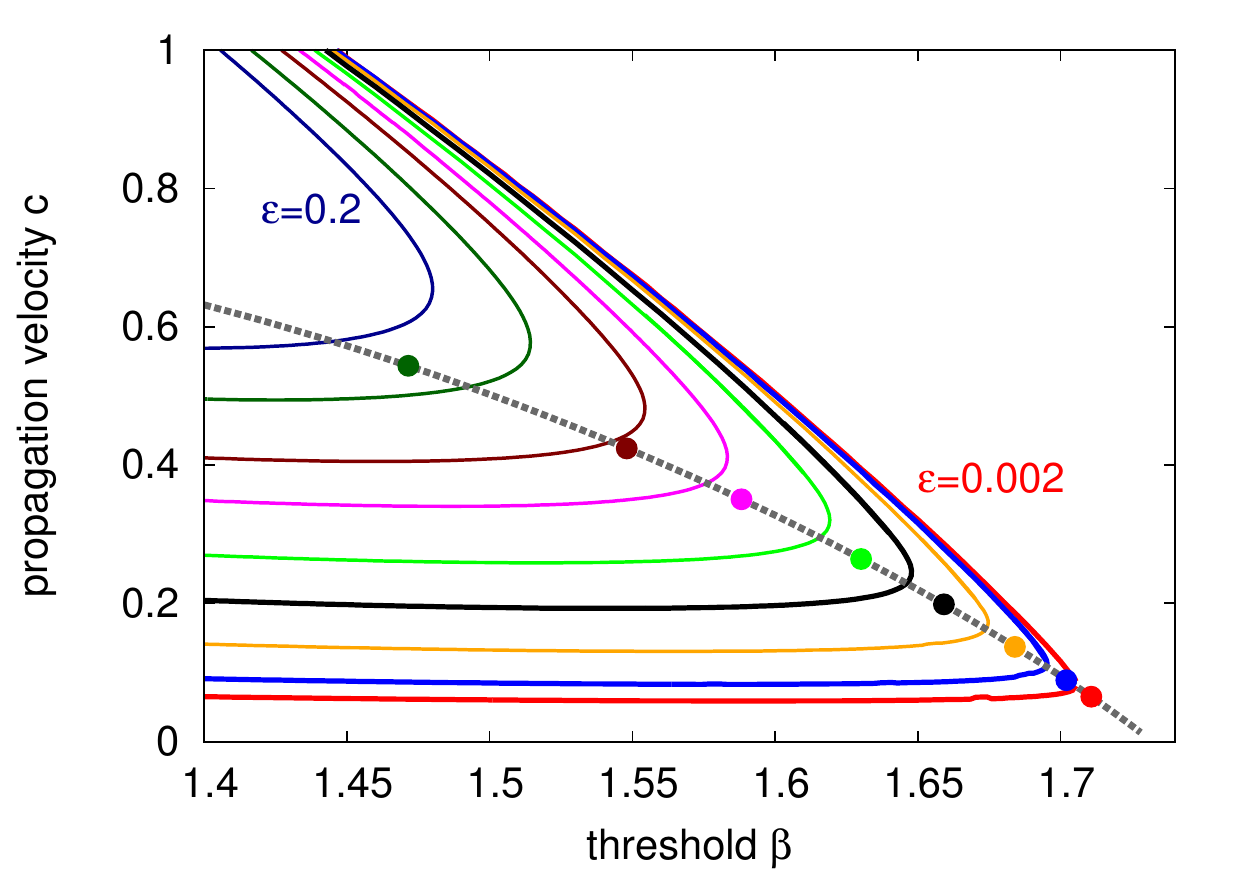}
\caption{Critical propagation velocity $c_{cr}|_{A=0}$ as a function of threshold $\beta$ (grey dashed line) derived from Eq.(\ref{eq:ccr}). The marked position (dots) on this line correspond to $\varepsilon=0.002, 0.004, 0.01, 0.022, 0.04, 0.07, 0.1, 0.15$.  The solid lines show the propagation velocity $c|_{A=0}$ of the fast and the slow wave as a function of threshold $\beta$ numerically computed from Eqs.~(\ref{eq:FHNohneAu})-(\ref{eq:FHNohneAv}) with. The color code indicates the same $\varepsilon$, $A=0$ in each case. Note that $\varepsilon=0.2$ is not in the co-domain of Eq.~(\ref{eq:evonb}),  see Fig.~\ref{fig:evonbmy}.}
  \label{fig:ccr}
\end{figure}

Fig.\ref{fig:ccr} shows the propagation velocity $c|_{A=0}$ as a function of threshold $\beta$. Numerical results obtained from Eqs.~(\ref{eq:FHNohneAu})-(\ref{eq:FHNohneAv}) show the branches of the fast wave and the slow wave for varying time scale ratio $\varepsilon$. The fast wave branch and the slow wave branch meet at a critical velocity $c_{cr}|_{A=0}$. The analytical expression for the critical velocity $c_{cr}|_{A=0}$ Eq.(\ref{eq:ccr}) is compared to the numerical results. The larger the threshold $\beta$ is, the better is the analytical approximation: For large threshold $\beta$ the saddle-node bifurcation, where the fast wave branch meets the slow wave branch, occurs for smaller time scale ratio $\varepsilon$, which in turn improves the results from the singular perturbation theory. The analytical results systematically lie below the numerical results, which is a consequence of the analytical result for the propagation velocity of the fast wave $c^f|_{A=0}$ being to small over a large range of 
parameters, see above.\\

\begin{figure}
\includegraphics[width=\columnwidth]{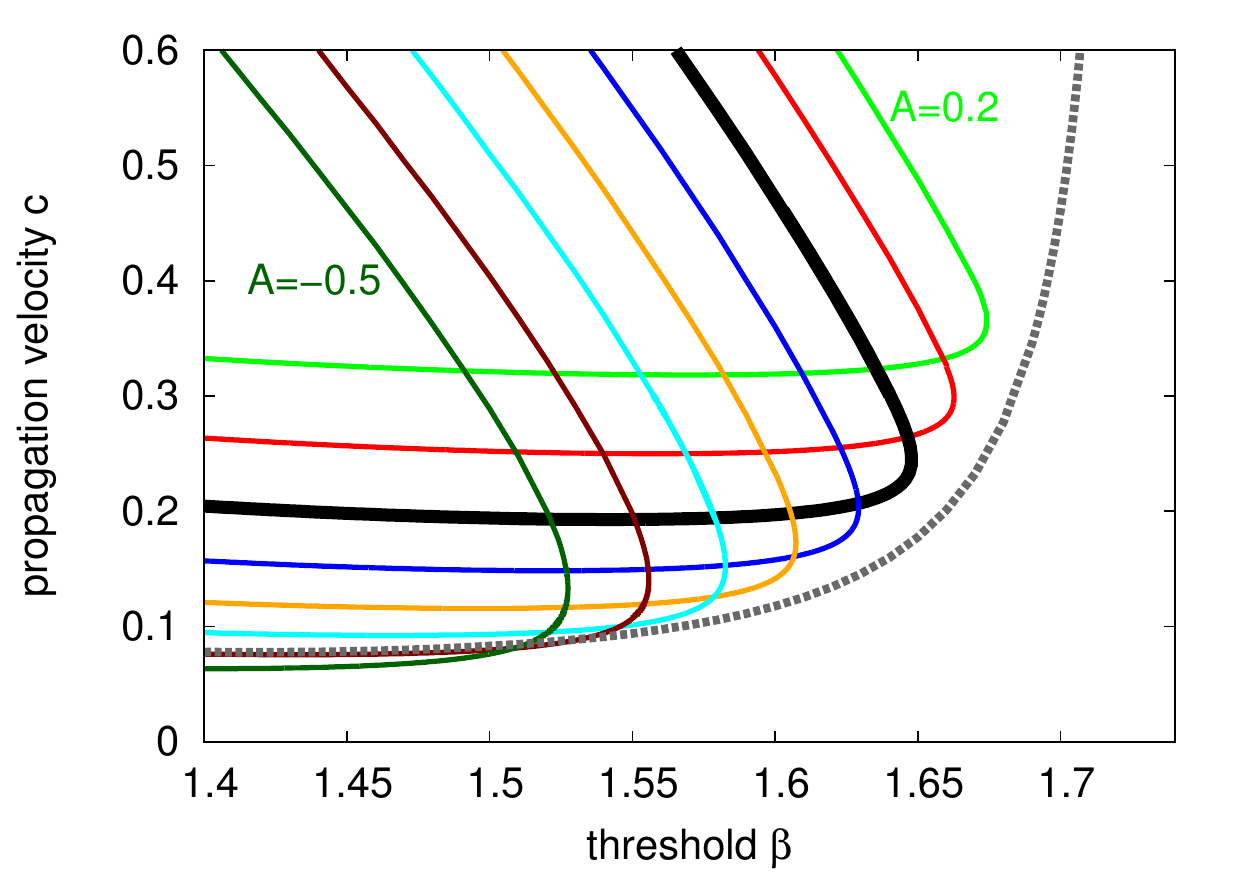}
 \caption{Critical propagation velocity $c_{cr}(A_{cr})$ as a function of threshold $\beta$ (grey dashed line) derived from Eq.(\ref{eq:cmin}) with $\varepsilon_{cr}$ from Eq.(\ref{eq:evonb}) and $c_{cr}|_{A=0}$ from Eq.(\ref{eq:ccr}). The coloured solid lines show the propagation velocity $c(A)$ of the fast and the slow wave numerically computed from Eqs.~(\ref{eq:FHNu})-(\ref{eq:FHNv}) with varying advection strength $A$ ($A=-0.5, -0.4, -0.3, -0.2, -0.1, 0.0, 0.1, 0.2$). $\varepsilon=0.022$ in each case.}
  \label{fig:cmin}
\end{figure}

Fig.\ref{fig:cmin} shows the propagation velocity of the fast, $c^f(A)$, and the slow, $c^s(A)$, wave affected by advection of varying strength $A$ as a function of threshold $\beta$. The branches of the fast and the slow wave velocity are numerically obtained from Eqs.~(\ref{eq:FHNu})-(\ref{eq:FHNv}). Also here, the fast and the slow wave branch meet at a critical velocity $c_{cr}(A)$. In addition, the analytical result for the critical velocity in the presence of advevction Eq.(\ref{eq:cmin}) is shown. It provides the same characteristic trend as the numerical results.  Referring to systems without advection, the propagation velocity $c(A)$ is decelerated for negative advection strength $A<0$. The propagation boundary $\partial P$ (connection between fast and slow wave branch) is shifted to smaller threshold $\beta$. Traveling waves affected by positive advection $A>0$ are accelerated, the propagation boundary $\partial P$ is shifted to larger threshold $\beta$.\\
A theoretical explanation of the stabilizing effect of postive advection has been found: every parameter point in the ($\varepsilon,\beta$) space can be allocated a critical velocity (Eq.(\ref{eq:cmin})). Media without advection are excitable, if the propagation velocity of the fast wave is larger than this critical velocity (parameter regime above the critical line in Fig.(\ref{fig:evonbmy})) and non-excitable, if the propagation velocity of the fast wave is smaller than this critical velocity (parameter regime below the critical line in Fig.(\ref{fig:evonbmy})). Negative advection $A<0$ causes a deceleration of traveling waves, which in turn can induce a destabilization of an originally stable wave, if the fast wave is decelerated below the critical velocity $c_{cr}(A_{cr})$ \citep{ ZYK84}. On the contrary, positive advection $A>0$ causes an acceleration of traveling waves, which in fact can induce stable wave propagation in the former non-excitable parameter regime, if the fast wave is accelerated above 
the 
critical velocity $c_{cr}(A_{cr})$.\\

\begin{figure}
\includegraphics[width=\columnwidth]{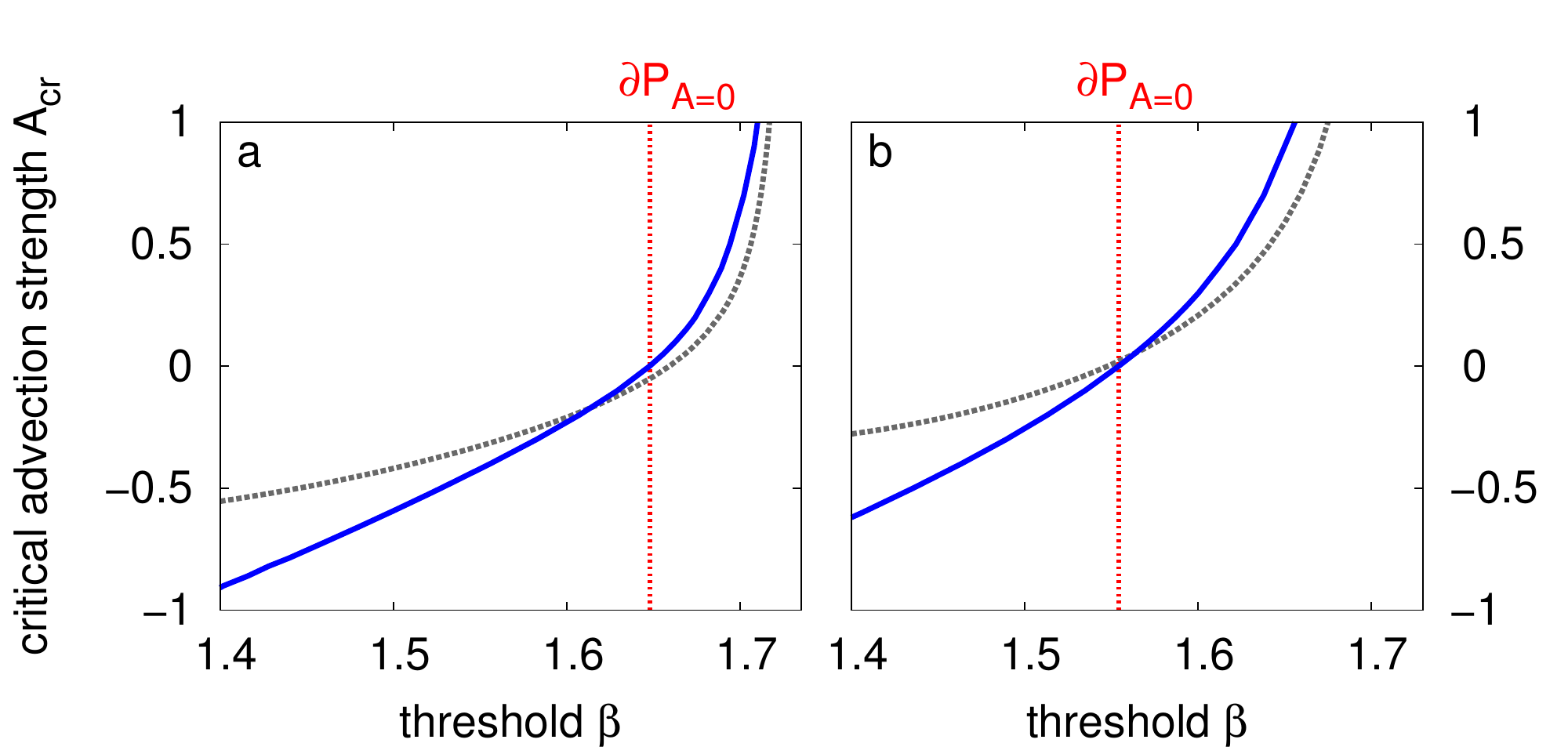}
 \caption{Critical advection strength $A_{cr}$ as a function of threshold $\beta$ for two different values of time scale ratio $\varepsilon$ (a) $\varepsilon=0.022$; b) $\varepsilon=0.1$). The grey dashed line shows the results derived from Eq.(\ref{eq:Acr}) with $\varepsilon_{cr}$ from Eq.(\ref{eq:evonb}) and $c_{cr}(A_{cr})$ from Eq.(\ref{eq:ccr}); the blue solid lines show the results numerically computed from Eqs.~(\ref{eq:FHNu})-(\ref{eq:FHNv}). The propagation boundary $\partial P_{A=0}$ is numerically computed from Eqs.~(\ref{eq:FHNohneAu})-(\ref{eq:FHNohneAv}).}
  \label{fig:Acr}
\end{figure}

\begin{figure}
\includegraphics[width=\columnwidth]{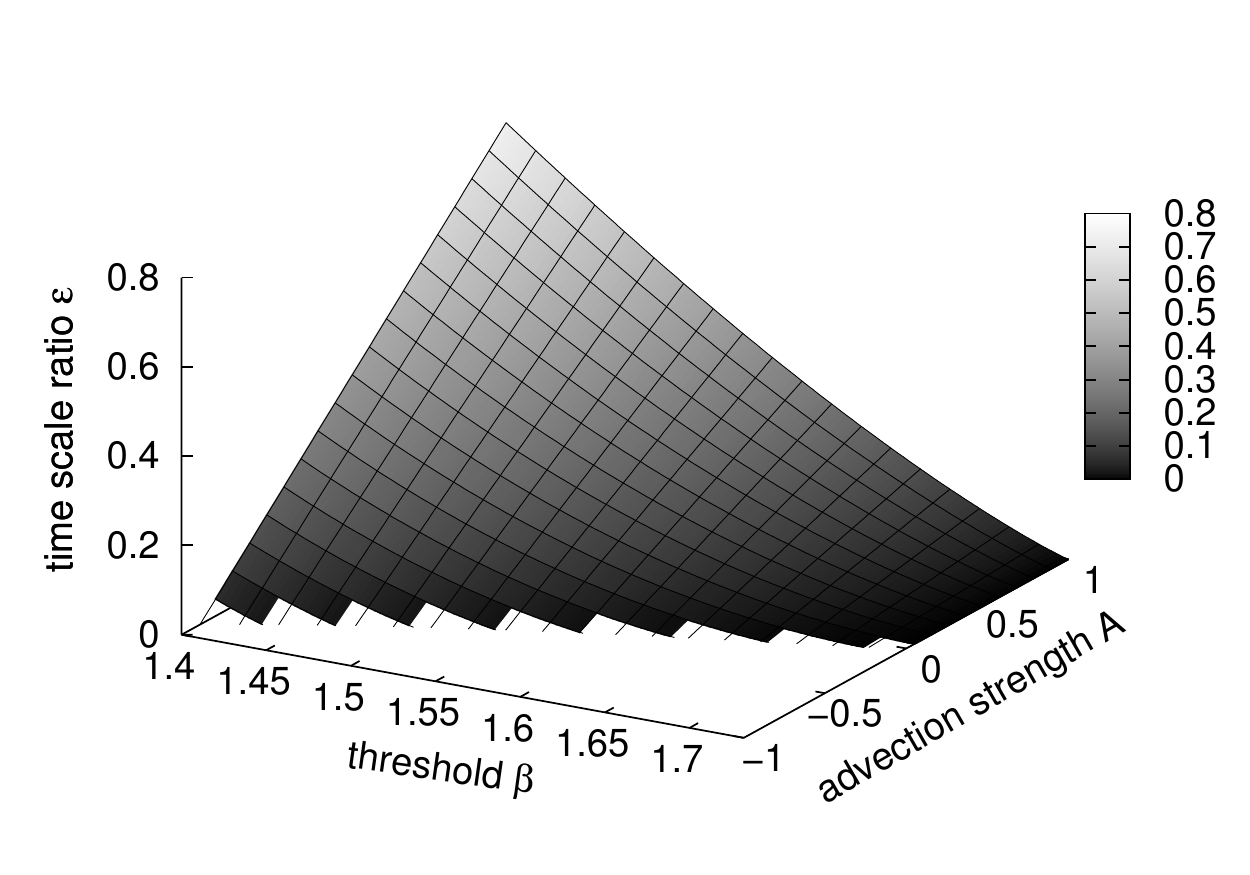}
 \caption{The critical surface in the $(\varepsilon,\beta,A)$ parameter space derived from Eq.(\ref{eq:criticalsurface}) separates the excitable (below) and the non-excitable (above) parameter regime.}
  \label{fig:crsurface}
\end{figure}

In Fig.(\ref{fig:Acr}), the critical advection strength $A_{cr}$ is shown in the $(\beta,A)$ parameter space for two different values of time scale ratio $\varepsilon$. The analytical results from Eq.(\ref{eq:Acr}) are compared to numerical results obtained from Eqs.~(\ref{eq:FHNu})-(\ref{eq:FHNv}). We find, that Eq.(\ref{eq:Acr}) provides the same characteristic trend as numerical results, but deviates strongly from numerical line for large negative adevction strength $A<0$, as in  this case $\varepsilon^*$ Eq.~(\ref{eq:epsstar}) is very large, and thus the singular perturbation theory breaks down. The critical line in the $(\beta,A)$ parameter space separates the excitable ($A>A_{cr}$) and the non-excitable ($A<A_{cr}$) parameter regime. Compared to systems without advection, the propagation boundary $\partial P$ is shifted to smaller threshold $\beta$ for negative advection $A<0$ and to larger threshold $\beta$ for positive advection $A>0$.\\

Fig.\ref{fig:crsurface} finally shows the critical surface in the $(\varepsilon,\beta,A)$ parameter space derived from Eq.(\ref{eq:criticalsurface}). It separates the excitable and the non-excitable parameter space, for values of $\varepsilon$ above the critical surface, propagating waves are not supported.\\

\section{Conclusion}
\label{sec:conclusion}
In this work, we described the dependency of the propagation velocity of an unstable slow traveling wave $c^s(A)$ on advection of strength $A$ analytically (Eq.(\ref{eq:cs12})) and numerically. We have shown, that positive advection \mbox{$A>0$}, corresponding to a constant field externally applied parallel to the propagation direction respectively corresponding to a small positive curvature (V-shaped pattern), can induce stable propagation of traveling waves in the non-excitable parameter regime. This behaviour is explained analytically: Every point in the $(\varepsilon,\beta)$ space, where $\varepsilon$ is the time scale ratio and $\beta$ is a measure for the threshold of the system, is related to a critical velocity $c_{cr}(A_{cr})$ (Eq.(\ref{eq:cmin})). $c_{cr}(A_{cr})$ is the propagation velocity at a saddle-node bifurcation of an unstable slow and a stable fast traveling wave solution, thus the minimal possible velocity of the fast wave solution. Stable wave propagation in the non-excitable parameter regime 
now 
is induced by accelerating the fast wave velocity above the critical velocity by affecting it with advection larger than a critical advection strength $A_{cr}$ (Eq.(\ref{eq:Acr})). We derived an analytical approximation of a critical surface in the $(\varepsilon,\beta,A)$ space (Eq.(\ref{eq:criticalsurface})), above which wave propagation is impossible.
Finally, we confirmed numerically, that the calculated dependencies of the critical velocity $c_{cr}(A_{cr}$) and the critical advection strength $A_{cr}$ on $\beta$ and $\varepsilon$ are valid in the in systems without advection excitable and non-excitable parameter regimes.

\section{Acknowledgments}
This work was supported  by the Bundesministerium f\"ur Bildung und Forschung (BMBF 01GQ1109) and by DFG in the framework of SFB 910.

\section{Appendix}
\label{sec:Appendix}

\subsection{Critical advection strength derived from nonlinear Eikonal equation}

The nonlinear Eikonal equation is given by (see Eq.(\ref{eq:cf12})) 
\begin{eqnarray}
 c^f_{\pm }(A)&=&\frac{1}{2}((A+c_0+\varepsilon c_1)\pm\sqrt{(A+c_0+\varepsilon c_1)^2-4\varepsilon A c_1}).\nonumber\\
\end{eqnarray}
The propagation velocity $c^f_+(A)$ remains real only, if the discriminant is larger than zero. Hence the limiting allowable advection strength $A_{cr}$ is determined by 
\begin{eqnarray}
 (A_{cr}+c_0+\varepsilon c_1)^2-4\varepsilon A_{cr} c_1&=&0.
 \label{eq:agleichnull}
\end{eqnarray}
Solving Eq.(\ref{eq:agleichnull}) for $A_{cr}$ yields
\begin{eqnarray}
 A_{cr}^{\pm}&=&-(c_0-\varepsilon c_1 \pm2\sqrt{-c_0\varepsilon c_1}).
 \label{eq:acreikonal}
\end{eqnarray}
The critical advection strength $A_{cr}$ is $A_{cr}^-$, because $|A_{cr}^+|>|A_{cr}^-|$.

\subsection{Expression for $c_0$, $c_1^f$ and $c_1^s$}

The exact analytical expression for the propagation velocity of the stable fast inner solution of FitzHugh-Nagumo model to lowest order of $\varepsilon$ is 
\begin{eqnarray}
 c_0&=&\sqrt{\frac{D}{2}}(u_1+u_3-2u_2),
\end{eqnarray}
with $u_1$, $u_2$ and $u_3$ being the intersection points of the $u$-nullcline with the inhibitor fixpoint $v_0=-3\beta+\beta^3$,
$u_1=-\beta$, $u_2=\frac{\beta}{2}-\sqrt{3-3/4\beta^2}$, and $u_3=\frac{\beta}{2}+\sqrt{3-3/4\beta^2}$.\\

The correction to first order of $\varepsilon$ of the propagation velocity of the inner stable fast wave solution considering solitary waves is
\begin{eqnarray}
 c_1^f&=&-\frac{\int_{-\infty}^{\infty}v_1\frac{\partial u_0}{\partial\xi} e^{-c_0\xi}\textit{d}\xi}{\int_{-\infty}^{\infty}(\frac{\partial u_0}{\partial\xi})^2 e^{-c_0\xi}\textit{d}\xi}
\end{eqnarray}
where $v_1$,
\begin{eqnarray}
 v_1(\xi)&=&\frac{1}{c_0}(u_3-u_1)(\xi+(\frac{\sqrt{2}}{u_3-u_1}\ln(1+e^{-\frac{u_3-u_1}{\sqrt{2}}\xi})),\nonumber\\
\end{eqnarray}
is the correction to first order of $\varepsilon$ of the inhibitor concentration (inner solution) of the fast wave and $u_0$, 
\begin{eqnarray}
 u_0(\xi)&=&\frac{u_1+u_3}{2}+\frac{u_1-u_3}{2}\tanh(\frac{1}{\sqrt{2}}\frac{u1-u3}{2}\xi),
\end{eqnarray}
is the (exact) inner solution of the activator concentration to order zero of $\varepsilon$.\\
The correction to order $\sqrt{\varepsilon}$ of the propagation velocity of the inner unstable slow wave solution is
\begin{eqnarray}
 c_1^s&=&\sqrt{\frac{2\sqrt{2m}-2l\ln\alpha}{\frac{(2m)^{(3/2)}}{3}-\frac{l^2}{2}\sqrt{2m}+\frac{l(l^2-2m)}{2}\ln\alpha}},
\end{eqnarray}
where $\alpha=\sqrt{\frac{l+\sqrt{2m}}{l-\sqrt{2m}}}$ and $l=\frac{2}{3}(-2u_1+u_2+u_3)$ and $m=(u_2-u_1)(u_3-u_1)$. For details, see Ref.\citep{Casten123456}.

\end{document}